\documentclass[conference]{IEEEtran}
\IEEEoverridecommandlockouts
\usepackage{cite}
\usepackage{amsmath,amssymb,amsfonts}
\usepackage{algorithmic}
\usepackage{graphicx}
\usepackage{textcomp}
\usepackage{xcolor}
\def\BibTeX{{\rm B\kern-.05em{\sc i\kern-.025em b}\kern-.08em
    T\kern-.1667em\lower.7ex\hbox{E}\kern-.125emX}}
\usepackage[bottom]{footmisc} 
\usepackage{parskip}

\usepackage{xspace}
\newcommand{\system}{CoverHunter\xspace}
\newcommand{\training}{coarse-to-fine training\xspace}

\usepackage{color}
\usepackage{multirow}
\usepackage{booktabs} 
\let\OLDthebibliography\thebibliography
\renewcommand\thebibliography[1]{
  \OLDthebibliography{#1}
  \setlength{\parskip}{0pt}
  \setlength{\itemsep}{0pt plus 0.3ex}
}

\begin{document}

\title{CoverHunter: Cover Song Identification with Refined Attention and Alignments}

\author{
\IEEEauthorblockN{Feng Liu}
\IEEEauthorblockA{\textit{
Intelligent Media Technology Department} \\
\textit{Huya Inc}\\
GuangZhou, China \\
liufeng1@huya.com} \\

\IEEEauthorblockN{Yinan Xu}
\IEEEauthorblockA{\textit{
Intelligent Media Technology Department} \\
\textit{Huya Inc}\\
GuangZhou, China \\
xuyinan@huya.com} \\

\and

\IEEEauthorblockN{Deyi Tuo}
\IEEEauthorblockA{\textit{
Intelligent Media Technology Department} \\
\textit{Huya Inc}\\
GuangZhou, China \\
tuodeyi@huya.com} \\


\IEEEauthorblockN{Xintong Han* \thanks{\hrule \vspace{3pt} *Xintong Han is the corresponding author}}
\IEEEauthorblockA{\textit{
Intelligent Media Technology Department} \\
\textit{Huya Inc}\\
GuangZhou, China \\
hanxintong@huya.com}
}

\maketitle

\begin{abstract}
Cover Song Identification (CSI) focuses on finding the same music with different versions in reference anchors given a query track. In this paper, we propose a novel system named \system that overcomes the shortcomings of existing detection schemes by exploring richer features with refined attention and alignments. 
\system contains three key modules: 1) A convolution-augmented transformer (e.g. Conformer) structure that captures both local and global feature interactions in contrast to previous methods mainly relying on convolutional neural networks; 2) An attention-based time pooling module that further exploits the attention in the time dimension; 3) A novel \training scheme that first trains a network to roughly align the song chunks and then refines the network by training on the aligned chunks. At the same time, we also summarize some important training tricks used in our system to achieve better results. Experiments on several standard CSI datasets show that our method significantly improves over state-of-the-art methods with an embedding size of 128 (2.3\% on SHS100K-TEST and 17.7\% on DaTacos). 
\end{abstract}


\section{Introduction}
Cover Song Identification (CSI) targets finding the same music work with different versions given a query track. In recent years, CSI has become increasingly important with the development of online music platforms and apps. In particular, the CSI algorithms have made outstanding contributions to detecting to detecting cover songs and protecting music copyrights. As a contrastive learning problem, the diversity of music (variants of keys, tempos, different instruments played, etc.) and the requirement for real-time performance challenge the algorithm.

In the early days of CSI development, handcrafted features were implemented to achieve decent results \cite{marolt2006mid,wang2003industrial,ellis2007identifyingcover}. However, these traditional methods have two main disadvantages. First, the accuracy remains unsatisfactory, especially with a lower recall, as handcraft features cannot handle various music styles and instruments. Second, the computational cost is too high to meet the real-time requirements of online applications due to the rapidly growing data scale.

Therefore, neural network methods have become mainstream and achieved promising progress in large data sets. The most popular paradigm is training a CNN-based model to extract version embedding by jointly minimizing classification loss and contrast loss. For example, Yu et al. proposed TPPNet (temporal pyramid pooling) \cite{yu2019temporal} and CNN-based CQTNet \cite{yu2020learning} for learning the characteristics of the cover song task. ByteCover \cite{du2021bytecover} and ByteCover2 \cite{du2022bytecover2} achieved state-of-the-art results with a ResNet-IBN50 \cite{seetharaman2017cover} backbone and multi-loss training with cross-entropy and triplet loss. PiCKINet \cite{o2021detecting} proposed Pitch Class Blocks in order to maintain the key-invariance features of music. LyraC-Net \cite{hu2022wideresnet} utilized WideResNet as the backbone, and combined classification and metric learning for optimization.


\begin{figure}[t]
  \centering
  \includegraphics[width=0.98\linewidth]{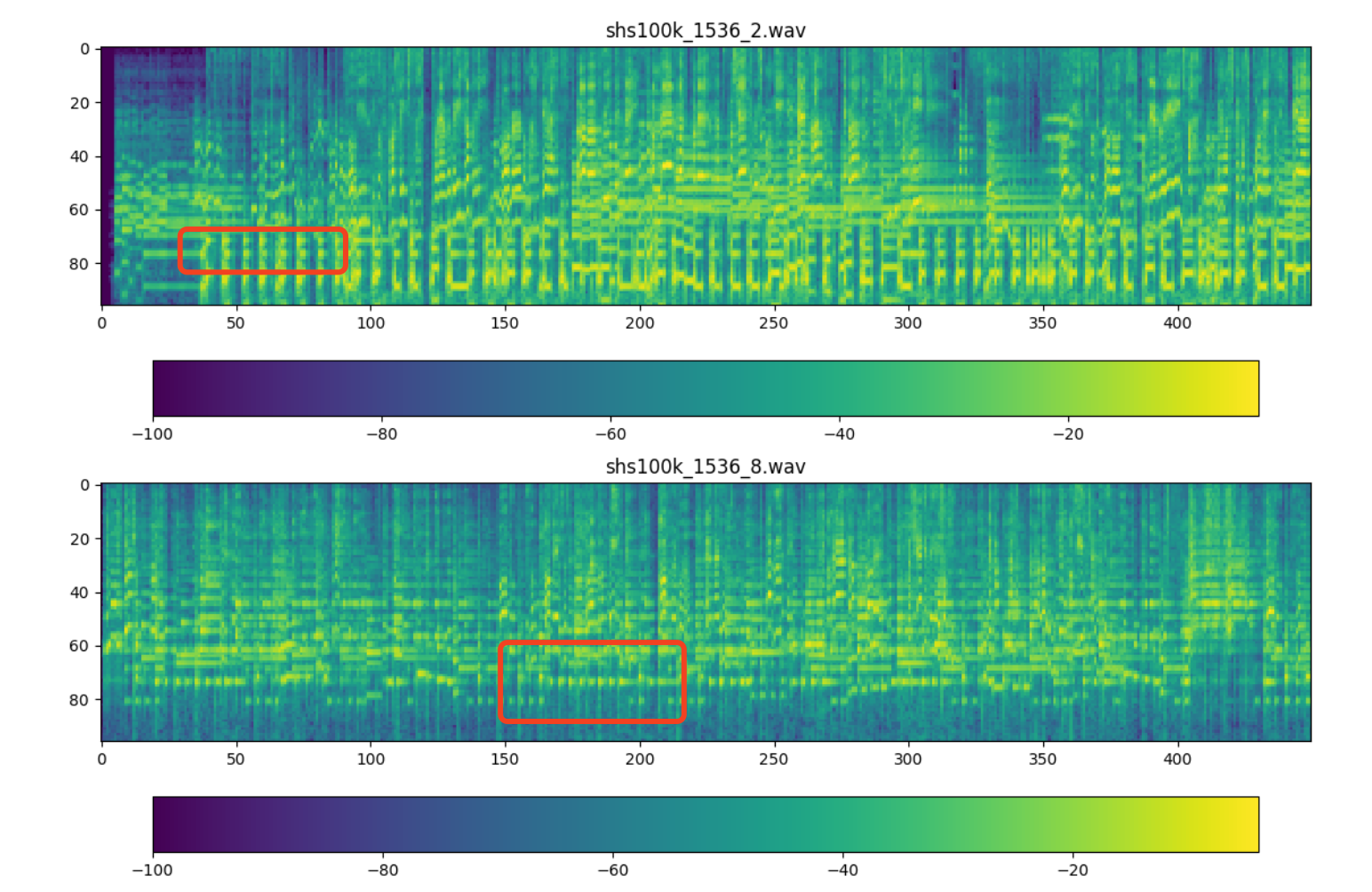}
  \caption{CQT features of two different versions of the song \textit{America the Beautiful}. The above is the version of BeBe Winans, and the below is from Keb' Mo'. Audio from SHS100K, song-1536, version-2/8.}
  \label{fig:Comparison_cqt}
\end{figure}

Although CNN-based features have significantly outperformed the traditional handcrafted features by experts, most existing methods directly utilize a metric learning framework in an end-to-end fashion while ignoring the underlying information specific to the CSI task. We posit that such CSI-related information could be better mined and utilized, significantly boosting the final performance.
For example, the CNN-based models typically focus on local information and neglect global information. 
As highlighted by the red boxes in Fig. \ref{fig:Comparison_cqt}, songs usually contain many periodic signals vital for detecting cover songs in addition to the features we typically think of (chords, beats, etc.). And these periodic signals typically span an extended range of time and thus cannot be modeled with a CNN-based backbone.

To overcome such shortcomings, we propose to use the Conformer \cite{gulati2020conformer} to capture global information with the self-attention mechanism while still accounting for local clues with convolutions. Moreover, when pooling the features across the time domain to obtain a fixed-length audio embedding, we abandon the popular choice of average pooling or generalized mean pooling \cite{du2021bytecover} that treats the features equivalently across time. Instead, we introduce a self-attention mechanism to amplify the contribution of discriminative frames and suppress redundant information that is unrelated to identifying cover songs. 

Another problem encountered in cover detection is that for two songs that are covers of each other, only part of the segment may be a cover, as cover songs often include some free-designed fragments by the creator at the beginning or end. For example, 17s in the upper example in Fig. \ref{fig:Comparison_cqt} corresponds to 29s in the lower song since the first 10s in the lower example is a user-defined prelude, which not matching any other version of the song. As a result, if we directly input these fragments that have nothing to do with the original song into the training data, it will inevitably degrade the performance.
In other words, unlike other contrastive learning problems (e.g. speaker recognition), the cover song detection has an alignment issue -- we need to learn informative features from audio with different lengths and a large number of redundant and unaligned clips.

To address the alignment challenge, we propose a novel \training, first training a coarse model on short song chunks to obtain representations to roughly align the training samples. Then, we feed the aligned longer chunks to train a refined model that produces more discriminative features without the interference of unaligned and unrelated audio content.

In sum, \system makes the following contributions.

$\bullet$ We propose a new Conformer-based model, which combines convolutions to extract local information and attention to extract global information. While Conformer \cite{gulati2020conformer} has achieved good results in audio speech recognition, as far as we know, we are the first to make it work for the CSI task. 

$\bullet$ We redesign the time domain pooling module with attention to effectively suppress redundant information and highlight key features.

$\bullet$ We propose a new \training mode in contrast to the traditional training procedure, obtaining more aligned features for learning the song similarity.

$\bullet$ We also summarize a bag of training tricks (e.g. data augmentation) that allow us to achieve better results and could be easily leveraged in other tasks in the community.

$\bullet$ \system outperforms other methods by a clear margin on standard datasets like SHS100K and DaTacos.

\section{Method}

Our system contains three main components: CQT-extractor, contrastive learning, and vector retrieval. In this section, we describe the contrastive leaning part in detail as summarized in Fig. \ref{fig:model_proposed}, which includes (1) a Conformer-based backbone (Section \ref{sec:conformer}), (2) an attention-based time domain pooling module (Section \ref{sec:time-pool}), (3) joint loss containing focal loss, center loss, and triplet loss (Section \ref{sec:loss}), and (4) a novel \training scheme for learning discriminative features with song chunk alignment (Section \ref{sec:train-align-train}).

\subsection{Conformer Backbone}
\label{sec:conformer}

As shown in Fig. \ref{fig:model_proposed}(b), the model takes as input the constant-Q transform (CQT) \cite{brown1991calculation} spectrum followed by a BatchNorm layer that replaces the traditional CMVN module to get normalized data. Then a trainable down-sampling module improves the training efficiency by reducing time-wise length by a factor of 4. We stack $N$ Conformer blocks on top of the down-sampled data, of which the structure is followed by the classic architecture used in speech recognition task \cite{gulati2020conformer}. We set $N=6$ and the number of channels in the attention layer to 256. A time-domain pooling module (Section \ref{sec:time-pool}) produces a fixed-length song embedding from frame embeddings.
Finally, with a simple linear bottleneck layer, we use the embeddings before and after the bottleneck to compute the corresponding losses (see Section \ref{sec:loss}).

\subsection{Time Domain Pooling Module}
\label{sec:time-pool}

The time domain pooling extracts a chunk-level embedding by aggregating frame-level embeddings. In our experiments, this pooling module also plays an important role in boosting the performance as we want to suppress the information in redundant frames and highlight the discriminative ones. To this end, we draw inspiration from the Attentive Statistics Pooling \cite{okabe2018attentive} and redesign it as illustrated shown in Fig. \ref{fig:model_proposed}(c). We first compute the mean and std of the input data on the time channel and concatenate them with the original feature. Then, we use self-attention to capture the global feature interactions, and the mean/std of the attended features are computed followed by a linear layer to reduce the data dimension.
In the self-attention layer, we use a random mask to erase some value of the embedding to improve the robustness further. 
As a result, the introduced time-pooling module leverages feature statistics and pays more attention to the discriminative frames, which leads to clear outperformance compared to the widely-adopted generalized mean pooling (GeM) in recent approaches \cite{du2021bytecover,du2022bytecover2,hu2022wideresnet}. More ablation studies will be shown in section \ref{sec:ablation_study}.

\subsection{Loss}
\label{sec:loss}

We use the embedding before the bottleneck layer to compute the contrast loss (i.e., triplet loss \cite{schroff2015facenet}) and use the embedding after the bottleneck layer to compute the focal loss \cite{lin2017focal} and center loss \cite{wen2016discriminative}. Note that we use the focal loss to replace the traditional cross-entropy loss as in \cite{du2021bytecover}, improving the performance when facing data unbalance. Center loss \cite{wen2016discriminative} helps the training convergence and achieves higher performance. The detailed loss functions are as below: 

\begin{align}
 &L_{focal} = -\alpha_i(1-p_i)^\gamma \log(p_i), \\
 &L_{center} = 0.5 * \left \| x_i - c_{y_i} \right\|^2_2, \\
 &L_{triplet} = \max(d_p -d_n + \alpha, 0),
\end{align}
where $L_{focal}$, $L_{center}$, $L_{triplet}$ denotes focal loss, center loss, and triplet loss, respectively. $p_i$ is the estimated probability of the $i$-th sample being correctly classified, and $\-\alpha_i$, which is inversely proportional to the number of samples in the class, weights each loss term. $c_{y_{i}}$ is the feature center of the class $y_i$. $d_p$ is the distance between a positive pair, while $d_n$ is that of a negative pair. We use the default parameters as in the original papers $\gamma=2$, $\alpha=0.3$. At last we train \system by minimizing $L_{total}$:
\begin{equation}
 L_{total} = \lambda_{focal} L_{focal} + \lambda_{center} L_{center} + \lambda_{tri} L_{triplet} 
\end{equation}
with $\lambda_{focal} = 1.0$, $\lambda_{center} = 0.01$, $\lambda_{tri} = 0.1$, balancing different loss terms.

\subsection{Coarse-to-fine Training Scheme}
\label{sec:train-align-train}

Accurate alignment of audio information is important to improve performance. To fully exploit the alignment information, we propose a coarse-to-fine training scheme with chunk-level alignment.

\noindent\textbf{Coarse training}. We first split the training samples into short fixed-length chunks (we use 15s short chunks with 7.5s overlap in our experiments) and train a coarse model using the track label as the label of its chunks. The embedding computed by the model will be further used to calculate the alignment information. 

\begin{figure*}[t]
  \centering
  \includegraphics[width=1\linewidth]{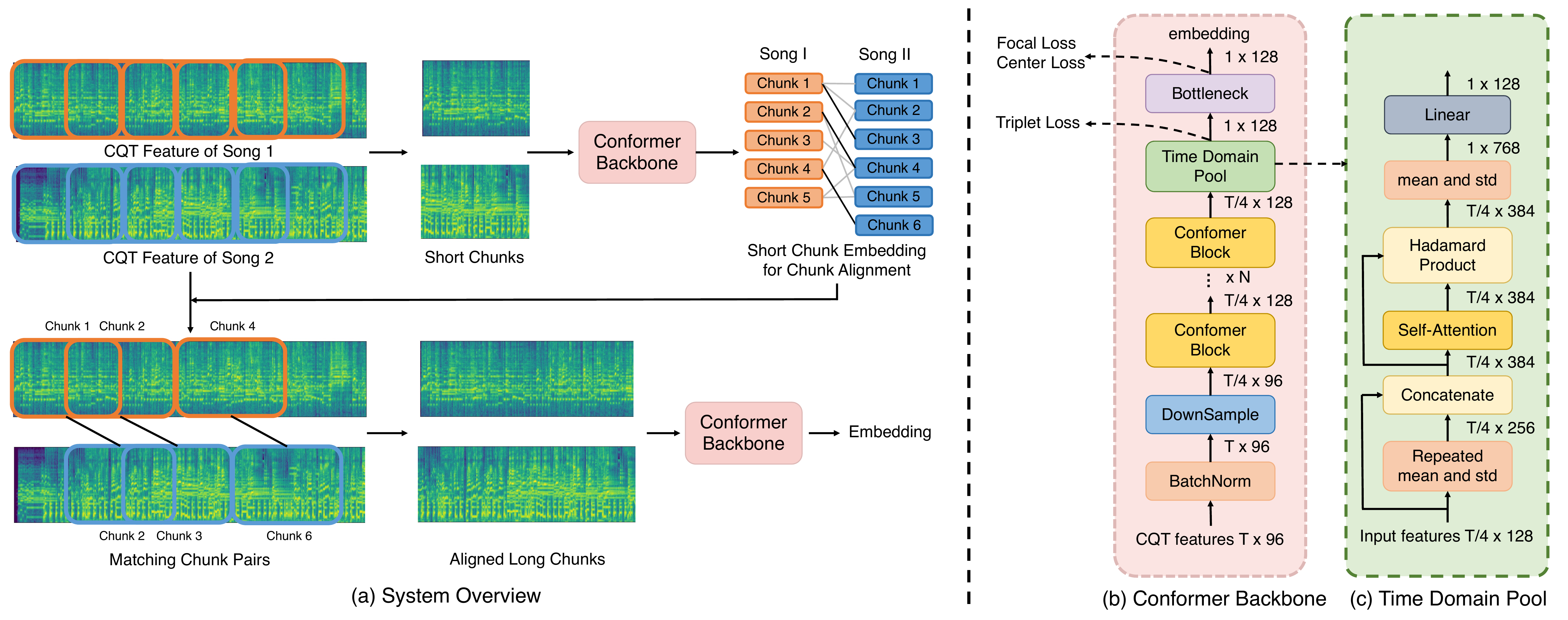}
  \caption{Overview of \system. (a) We first split the songs into 15s short chunks to learn short chunk embeddings for alignment. Then the aligned long chunks are used to train a refined model for cover song retrieval. (b) Conformer-based backbone with a time domain pooling module. (c) The detailed structure of our attention-based time domain pooling module.}
  \label{fig:model_proposed}
\end{figure*}

\noindent \textbf{Finding aligned chunks.} In this step, we aim to find aligned chunks of two covers with the same label. As shown in Fig. \ref{fig:model_proposed}(a), we compute the pairwise distance between these short chunk embeddings computed by the coarse model. 
The pairs with cosine similarity larger than a specified threshold (0.9 in our implementation) are selected as matching pairs. Here $(p_{i1}, p_{i2}, \delta_{i})$ denote the positions of two chunks in their covers and the relative offset $\delta_{i}=p_{i2}-p_{i1}$ of each matching pair $i$.
We get the mode (number occurring most) of all $\delta_i$ as $\Delta$. And we select all matching pairs with $\delta_i = \Delta$ as the aligned pairs. The start time of the aligned pairs is recorded to be used in the next step.
As for the example shown in Fig. \ref{fig:model_proposed}(a), we can get the following matching pairs (connected with lines in the figure):
\begin{align*}
&(1, 1, 0), (1, 2, 1), (1, 3, 2), (2, 4, 2), (2, 5, 3) \\ 
&(3, 4, 1), (4, 2, -2), (4, 6, 2), (5, 4, -1), (5, 5, 0) 
\end{align*}
and we get $\Delta=2$. Thus, $(1,3,2), (2,4,2), (4,6,2)$ are selected as aligned chunk pairs (denoted with black lines in the figure).

\noindent \textbf{Fine training.} For any aligned pair obtained from the last step, we extend the chunk to a longer length (e.g. 15s$\sim$45s) from the aligned start time. We then use these aligned pairs as positives and other sample pairs as negatives to train a fine model with the same network structure and loss function.

\noindent \textbf{Inference.} For inference, we cut the gallery audio to 45s non-overlapping chunks and use the fine model to extract gallery embeddings. Given a music query, we obtain its 45s non-overlapping chunk embedding and use the closet chunk pair between the query and gallery chunks to decide the label of the query.

\begin{table*}[h]
	\caption{
	Performance of different models for SHS100K-TEST, Covers80, and DaTacos. The best and second best results are denoted by \textbf{bold} and \underline{underline}, respectively. \\
	}
	\label{tab:Performance}
	\centering
  \begin{tabular}{lcccccccccccc}
    \toprule
        \multirow{2}{*}{Method} & \multirow{2}{*}{Dims↓} & \multicolumn{2}{c}{SHS100K-TEST \cite{xu2018key}} && \multicolumn{2}{c}{Covers80 \cite{ellis2007identifyingcover}}  && \multicolumn{2}{c}{DaTacos \cite{yesiler2019tacos}}\\
        \cline{3-4}\cline{6-7}\cline{9-10} &&  mAP$\uparrow$ & MR1$\downarrow$ &&  mAP$\uparrow$ & MR1$\downarrow$ && mAP$\uparrow$ & MR1$\downarrow$ \\
	\midrule        
        CQT-NET \cite{yu2020learning} & 300 & 0.655  & 54.9 && 0.840  & 3.85 && -  & - \\
        ByteCover \cite{du2021bytecover}  & 2048 & 0.836 & 47.3&&   0.906 & 3.54 && 0.714 & 23.0 \\
        ByteCover2-128 \cite{du2022bytecover2} & 128 & 0.839 & 45.5 && 0.912 & 3.40 && 0.718 & 22.7 \\
        ByteCover2-1536 \cite{du2022bytecover2} & 1536 & \underline{0.864} & 39.0 && 0.928 & 3.23 && 0.791 & 19.2 \\
        LyraC-Net \cite{hu2022wideresnet} & 1024 & 0.765 & 48.3 && \textbf{0.961} & \textbf{3.12} && 0.813 & 15.0 \\
        \system-128 & 128 & 0.858 & \underline{11.9} && 0.926 & 3.78 && \underline{0.845} & \underline{12.2} \\        
        \system-256 & 256 & \textbf{0.875} & \textbf{9.9} && \underline{0.933} & \underline{3.20} && \textbf{0.865} & \textbf{11.0} \\                
	\bottomrule
	\end{tabular}
\end{table*}

\section{Experiments}

\subsection{Experiment Settings}
\label{sec:dataset_and_settings}

\noindent \textbf{Datasets.} To fairly compare our method with others, we follow \cite{du2021bytecover,hu2022wideresnet} and use SHS100K \cite{xu2018key} as the training dataset. We download the data with URLs from second-hand-songs website\footnote{https://secondhandsongs.com/} and use the original setting\footnote{https://github.com/NovaFrost/SHS100K} to split all recordings into train/dev/test sets. In addition to evaluating our method on the test part of SHS100K, we also use Cover80 \cite{ellis2007identifyingcover} and DaTacos \cite{yesiler2019tacos} as test sets for comparison, which have 160 and 15,000 recordings, respectively. Note that we remove the overlap items between SHS100K and DaTacos as suggested by \cite{du2022bytecover2}.

\noindent \textbf{Evaluation metrics.} We follow the evaluation protocol of the MIREX Audio Cover Song Identification Contest\footnote{www.music-mir.org/mirex/wiki/2020:Audio\_Cover\_Song\_Identification} and use the mean average precision (mAP) and the mean rank of the first correctly identified cover (MR1) for evaluation.

\noindent \textbf{Implementation details.} In the CQT feature extractor, we set the     CQT bin size as 96. We use the hop size/length of 0.04s/0.08s, which results in 25 frames per second. We implement our model with the PyTorch framework. The model is trained using Adam Optimizer with the default hyper-parameters. The learning rate is set to 0.001 and decayed with a factor of 0.95 every 1,000 steps. We train our model with a batch size of 256 on a single NVIDIA V100 GPU. During inference, we use the cosine distance between two embedding vectors to measure the similarity between two musical chunks. We search for the most similar 45s chunk pair between the query and gallery chunks. By using vector retrieval methods such as ScaNN \cite{avq2020}, the search efficiency can be improved greatly. 

\subsection{Training Tricks}
\label{sec:training_tricks}

Before diving into the experiments, we summarize some training tricks which effectively improve performance.

\noindent \textbf{Noise mixing.} We mix suitable noise to the song, which mimics the real-world scenario. We choose a noise from Musan \cite{snyder2015musan} with a random SNR from -10db to 30db and mix it with a training sample. Noise mixing is especially effective for some noisy cases, such as in the DaTacos dataset.

\noindent \textbf{Data augmentation.} Data augmentation can improve the robustness of the model. We implement four types of augmentations: volume, speed, pitch, and mask augmentation. For volume, we randomly change the signal volume from -6db to 0db. For speed augmentation, we change the audio's speed with the pitch fixed. For pitch, CQT features allow us to roll the features along the frequency axis to change the pitch. For the mask augmentation, we follow \cite{luo2019bag} to randomly mask rectangle areas in the spectrum.

\noindent \textbf{Augmentation during training.} Instead of extracting all audio features before training, we do that on-the-fly during training after applying the aforementioned augmentations to the audio. As a result, the model sees the same sample with different augmentations for different epochs, which increases the data diversity and thus obtains a higher accuracy.

\noindent \textbf{Multi-length training.} Both CNN-based and attention-based methods can take audio with variable lengths as input. And we find it improves the model's robustness and generalizability to use samples with different lengths as input during training. For example, in this paper, we send aligned chunks with a random length from 15s to 45s at the training stage, and during inference, we use a fixed inference chunk length of 45s.

As presented in the ablation study (section 3.4), these tricks improve the vanilla baseline by 28.3\% (0.564 $\rightarrow$ 0.724) on DaTacos dataset. We believe these tricks will receive a wide range of adoption in CSI and related tasks in the future.

\subsection{Comparison on Performance}

We implement \system with embedding sizes of 128 and 256 and compare \system with state-of-the-art cover song identification methods CQT-NET \cite{yu2020learning}, ByteCover \cite{du2021bytecover}, ByteCover2 \cite{du2022bytecover2}, and LyraC-Net \cite{hu2022wideresnet} on the three datasets described in \ref{sec:dataset_and_settings}. The results are shown in Table \ref{tab:Performance}. We can see that the \system achieves higher mAPs and lower MR1s than other methods with the same dimension of embedding size (e.g. \system-128 vs ByteCover2-128).
Notably, \system sets a new state-of-the-art performance on SHS100K-TEST and DaTacos datasets with more compact representations (\system: 256 dims, ByteCover2: 1536 dims, and LyraC-Net: 1024 dims). 
As for the smaller-scale Covers80 dataset, our method obtains the second-best results with only a quarter embedding size compared to the best-performing model (e.g. 256 vs 1024).

\begin{figure}[t]
  \centering
  \includegraphics[width=1.0\linewidth]{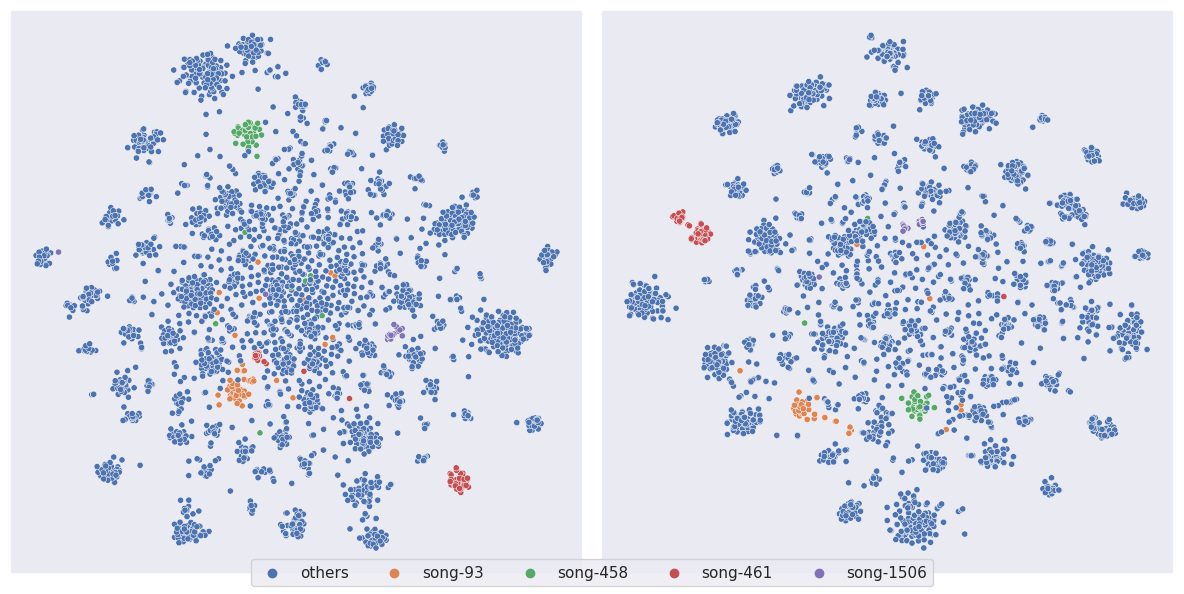}
  \caption{Embedding space of SHS100K-TEST: (a) \system without \training scheme. (b) \system with \training scheme.}
  \label{fig:tsne_for_embedding}
\end{figure}

At the same time, as we observe in our experiments, the mAP and MR1 of a given method do not have very strong correlations.
For example, on the SHS100K-TEST dataset, we get a relative improvement of MR1 much higher than mAP. 
This is because unlike mAP, which considers all sorted gallery audio tracks, MR1 only pays attention to the position of the first recalled result and does not consider other lower-ranked recalled gallery tracks. 
In practical applications (e.g. copyright infringement detection), another indicator we care about is the hit accuracy of the first recall result. To this end, we also use the hit rate metric to show the effectiveness of our method.
We get a hit rate of 0.931/0.901/0.935 on SHS100k-TEST/Covers80/DaTacos. For our \system, we find it more favorable when looking at the hit rate and MR1 metrics since we consider more global interaction and use aligned features for retrieval, which allows us to accurately find the most matched chunk and get the right top-1 result.

To better understand the advantage of our proposed model, we perform a visualization of the learned embedding with t-SNE \cite{van2008visualizing} in Fig \ref{fig:tsne_for_embedding}, which shows the distributions of embedding vectors without and with the \training scheme. We highlight several representative songs with different colors (i.e. points with the same colors represent different versions of the same song). 
The green samples are one of the relatively easy songs. Whether using the \training scheme or not, most of these samples are very close in the embedding space. Some hard samples (e.g. samples containing many irrelevant contents or facing large misalignment issues) cannot be correctly clustered without our \training (red and purple samples in Fig. \ref{fig:tsne_for_embedding}(a)), and they are pulled closer in the embedding space with \training scheme (Fig. \ref{fig:tsne_for_embedding}(b)). Interestingly, we find that sometimes covers of the same song are grouped into multiple cliques after \training, we attribute this to the reason that using  the attention mechanism and aligned chunks learns more refined and discriminative features such that it automatically discovers sub-classes (e.g. different types tempos or instruments played) for the same song.
The orange samples represent extremely hard cases, in which many outliers are present in Fig. \ref{fig:tsne_for_embedding}(a), although they are more tightly clustered after \training. Overall, the model with the \training scheme achieves better results because of the proposed alignment process for learning more informative features at a coarse-to-fine chunk level. 

\begin{table}[t]
	\caption{Ablation Study of \system-128 on DaTacos.}
	\label{tab:ablation}
	\centering
	\begin{tabular}{lcc}
		\toprule
		    Method  & mAP↑ & diff \\
		\midrule
        ResNet-IBN 50 (baseline)  & 0.724 & - \\
        ResNet-IBN 50 w/o training tricks & 0.564 & -21.7\% \\
        \midrule
        Conformer &  0.779 & 7.50\% \\
        Conformer + GeM pooling & 0.787 & 8.7\% \\
        Conformer + time-domain pooling & 0.806 & 11.3\% \\
        Conformer + \training & 0.832 & 14.9\% \\
        \system (Ours Full) & 0.845 & 16.7\% \\
		\bottomrule
	\end{tabular}
\end{table}

\subsection{Ablation Study}
\label{sec:ablation_study}

Table \ref{tab:ablation} presents an ablation study of key components in \system. For our baseline model, we choose ResNet-IBN 50\cite{pan2018IBN-Net} as the backbone following \cite{du2021bytecover,du2022bytecover2}. Unlike Bytecover, we don't use gem Pool as a base model, and data augmentation may be not exactly the same because of missing details. 
If we drop the training tricks described in section \ref{sec:training_tricks}, the mAP will significantly decrease by 21.7\%, validating the importance of the designed best practice of training a CSI model. Additionally, we explore the impact of the proposed modules on the results. Using Conformer to replace ResNet-IBN 50 (baseline) captures both the local information and the long-range relationship in the audio, which brings an improvement of 7.5\%. With the attention-based time pooling module, we further improve the mAP by about 3.8\%. The newly proposed \training scheme yields around 7.4\% mAP, validating that using aligned chunks to train the network mitigates the challenge of learning discriminative features from unaligned and redundant inputs. Finally, combining the \training scheme with our attention-based pooling module achieves the best performance (16.7\% better than the baseline). 

\subsection{Efficiency}

\setlength{\tabcolsep}{4pt}
\begin{table}[t]
\begin{center}
\caption{Time cost (ms) for a gallery with  one million songs.} 
\label{tab:efficiency}
\begin{tabular}{lcccc}
  \toprule
  Method & preprocess & inference & retrieval & total \\
  \midrule
  ResNet-IBN 50 & 192  & 108 & 70 & 370\\
  \system-128 & 192  & 54  & 70 & 316 \\
  \bottomrule    
\end{tabular}
\end{center}
\end{table}

Table \ref{tab:efficiency} illustrates the efficiency of our model. All experiments are conducted with a two-core Intel(R) Xeon(R) Platinum 8369B CPU @ 2.90GHz. We report the time cost of the three steps in \system-128 separately. Note that for different models, the CQT preprocess is shared, and the retrieval step takes the same time when the embedding dims are the same. \system is slightly faster than the widely-used CNN-based model (i.e. ResNet-IBN 50 in \cite{du2021bytecover,du2022bytecover2}). And for the retrieval part, we use the open-source ScaNN to search from a database containing one million music tracks.

\section{Conclusion}

By rethinking and exploring the nature of the cover song identification problem, we propose \system, a new system built on top of a Conformer-based backbone with an attentive time pooling module and a \training scheme to tackle this problem. The experiments show that we outperform current published baseline systems on several datasets, demonstrating the effectiveness of \system.
We expect more robust and accurate chunk alignment mechanisms to improve performance in future research. Moreover, we are deploying this method in industrial applications and solving practical problems in complex real-world scenarios.


\end{document}